\documentclass[]{aa}

\usepackage{txfonts}
\usepackage{graphicx}

\newcommand{\astorbdate}{Feb 20, 2007}

\begin{document}

   \title{Asteroid occultations today and tomorrow: \\ toward the GAIA era}

   \author{Paolo Tanga\inst{*}\inst{1}
    \and Marco Delbo\inst{1}
%    \inst{3}
    }

\offprints{P. Tanga}

\institute{\inst{1} Observatoire de la C\^ote d'Azur, BP4229, 06304 Nice Cedex 4\\
   \inst{*} \email{paolo.tanga@oca.eu}
%   \and Observatoire de la C\^ote d'Azur, BP4229, 06304 Nice Cedex 4\\
%   \email{Marco.Delbo@oca.eu}
%   \and ESA
   }

%   \date{\today; Received ; accepted}

\abstract {Observation of star occultations is a powerful tool
to determine shapes and sizes of asteroids. This is key
information necessary for studying the evolution of the asteroid belt
and to calibrate indirect methods of size determination,
such as the models used to analyze thermal infrared observations.
Up to now, the observation of asteroid occultations is an activity
essentially secured by amateur astronomers equipped with small,
portable equipments. However, the accuracy of the available
ephemeris prevents accurate predictions of the occultation events
for objects smaller than $\sim$100 km.}
{We investigate current limits in predictability and observability
of asteroid occultations, and we study their possible evolution in
the future, when high accuracy asteroid orbits and star positions
(such as those expected from the mission Gaia of the European                                       % MDB Eruopean -> European
Space Agency) will be available.}
{We use a simple model for asteroid ephemeris uncertainties and numerical
algorithms for estimating the limits imposed by the instruments, assuming realistic CCD
performances and asteroid size distribution, to estimate the expected
occultation rate under different conditions.}
{We show that high accuracy ephemerides which will be available in the
future will extend toward much smaller asteroids the
possibility of observing asteroid occultations, greatly
increasing the number of events and objects involved. A complete
set of size measurements down to $\sim$10~km main belt asteroids
could be obtained in a few years, provided that a small network of
ground-based 1m telescopes are devoted to occultation
studies.}{}


   \titlerunning{Asteroid occultations today and tomorrow.}

   \maketitle
%
%________________________________________________________________

\section{Introduction}
An accurate knowledge of asteroid sizes and size distribution is a
fundamental prerequisite for studying the origin and the
collisional evolution of these bodies, including the formation and
the dispersion of dynamical families. It also has important
implications for our understating of the mechanisms that led to
the formation of the terrestrial planets of our Solar System.
Furthermore, knowledge of the size of an asteroid allows its bulk
density to be inferred when the mass of the body can be independently
determined, for instance from the orbital period of an asteroid
satellite. Information about the density of asteroids is crucial
to study their interiors, and to constrain the
collisional processes leading to the formation and evolution of
the different populations of these minor bodies (e.g.
\cite{Bri02}).
Moreover, if the size of an asteroid is known, the albedo of the
body can be derived using Eq. \ref{H2D} (e.g. \cite{Bowell_et_al_AST2}), 
where $p_V$ is the geometric visible
albedo and $H$ is the absolute magnitude of the asteroid.
\begin{equation}\label{H2D}
 D(km)=\frac{1329}{\sqrt{p_V}} \times 10^{-H/5}
\end{equation}
Note that H is a quantity that can be easily obtained
from measurements of the apparent brightness of the object, for
example by means of photometric observations in the visible light.
The distribution of albedos and its correlation with the taxonomic
types and the orbital elements of asteroids is a key piece of data in
understanding their nature and their origin, and can help to
constrain the mineralogical composition of their surfaces.
Albedos are also crucial for developing and improving reliable and
truly debiased asteroid population models (e.g. \cite{SAMI},
\cite{STUBI04}).
%
%It is also essential, of course,
%for the determination of albedo and density. These two quantities
%are a precious key to access some knowledge about the physical
%properties of surfaces and interiors.
%
%MDB

Asteroid flybys of space probes have provided not only the best
``direct'' size determination, but also a wealth of information on
shapes and surface properties for a very limited number of these
bodies. However, a statistically significant sample of asteroid
sizes can be obtained only by remote observations.
Up to now, mainly indirect methods have been used to determine
asteroid diameters: the vast majority of asteroid sizes and
albedos have been derived from observations obtained in the
thermal infrared ($\sim$ 12 - 100 $\mu m$) by the Infrared
Astronomical Satellite (IRAS; \cite{IMPS, SIMPS}). However, it is
known that IRAS diameters of asteroids are not exempt of
systematic errors. They were derived by means of the Standard
Thermal Model (STM of \cite{theSTM}), whose results are dependent
on the value of a model parameter called the \emph{beaming factor}
and indicated with $\eta$. The value of $\eta$ was calibrated by
forcing the STM to give the correct diameters (as derived from
occultation observations) of the asteroids 1 Ceres and 2 Pallas
(\cite{Lebo86}). Several studies have highlighted, though, that a
value of $\eta$ appropriate for large main belt asteroids, might
lead to an underestimation of the true size when applied to small
asteroids whose surface thermal characteristics are likely
to be different from those of larger bodies (e.g. \cite{Spencer89,
MDB&AWH02, HarrLag02}).

Calibration of asteroid radiometric diameters was possible
because, in the past half century, the sizes of the largest
asteroids were also derived by measuring the duration of star
occultations. This approach has traditionally been flawed
by the poor accuracy of the prediction of the asteroid shadow path
on Earth, thus keeping the success rate of asteroid occultation
very low. The availability of precise star astrometry obtained by
the mission Hipparcos of the European Space Agency (ESA), has
improved the situation in the last $\sim$10 years
(\cite{Dunham02}), strongly increasing the number of successfully
measured asteroid occultation chords. 
Electronic equipments (fast video cameras) have also increased the
accuracy of occultation timings, previously obtained mainly by
visual observation. However, asteroid occultation observations
have remained mainly an activity reserved to observers equipped
with small, movable telescopes and equipment, ready to accept a
rather high rate of ``miss''.

In this study, after having reviewed the present state of asteroid
occultation studies, we will focus on a very likely evolution of
this observational approach, thanks to the future availability of
asteroid orbits and star positions whose uncertainties will be
reduced by orders of magnitude. In this respect, the astrometric
ESA mission {\sl Gaia} (\cite{Perry01}) will represent a major
milestone that will completely change the approach to asteroid
occultation studies, opening the way to the possibility of
obtaining a more complete mapping of asteroid sizes and albedos.
In this paper we address this forthcoming improvement,
exploring the path that the approach to asteroid occultations will
take in the future and investigating the limits that future
observation will be able to approach.

Note that indirect methods of asteroid size and albedo
determination, such as those based on polarimetry and thermal
infrared observations, will greatly benefit from occultation studies
performed on a larger sample of asteroids: it will be possible,
for example, to calibrate models used to analyze thermal infrared
and visual polarimetric measurements on a larger and more accurate
sample of calibration asteroid diameters, with object sizes
extending down to some tens of kilometers. %(\cite{Shevchenko06}).

Since better performing telescopes will be needed to push size
determination toward smaller and thus fainter asteroids, in this
paper, we also study the observing rate of asteroid
occultations expected for a fixed observing station using a
one-meter-class telescope equipped with a cooled CCD camera
capable of a high frame rate.

After having assessed the present state of occultation studies
(Sect. 1) we analyze the theoretical occultation rate that
is expected for different asteroid sizes (Sect. 2). We then
study the role of instrument detection capabilities (Sect. 3)
and that of ephemeris accuracy (Section 4). The implications of
high-accuracy ephemeris are then discussed (Section 5).

\section{The observation of asteroid occultation: where we are now}

\subsection{Observational campaigns}

The first occultation ever recorded is that of 3 Juno, in 1958.
The third one (433 Eros), observed in 1975, was also the first to be
observed by multiple sites, allowing the
determination of different ``chords'' of the asteroid (Millis and
Dunham 1989). Only 17 positive occultation observations were obtained by the
end of 1980.

Coordinated campaigns and last-minute small-field astrometry,
aimed both at improving the occultation path and at
diffusing the relevant predictions among observer networks, have
provided a moderate increase in the rate of successfully measured
asteroid occultations events. Up to the early 1990s the
probability of observing an occultation chord for a given event
was discouraging. However, consistent efforts were rewarded
with a total of 60 positive events by 1985, growing to 101 in
1990, and increasing more rapidly later, thanks to the
availability of more precise star astrometry. After the
publication of the Hipparcos and Tycho-2 star catalogues
(\cite{Hog2000}) the number of successfully recorded occultation
events doubled in the decade 1991-2000 (see below for an
evaluation of the current prediction accuracy). At the end of 2004
they summed up to 658, and today they continue to grow at a rate
of about $\sim$100 per year (\cite{Dunham05}).

Despite the apparently large number of observed events, the
extremely limited number of chords for most of the observed events
results in diameters that remain poorly constrained. Also, the
poor knowledge of shapes, rotational poles, and rotational phases
at the epoch of the events often makes the interpretation of
observations difficult. In practice, the use of occultation
diameters found in current datasets\footnote{Interested readers
should refere to the PDS asteroid occultation dataset (AOD), available
from the NASA Planetary Data System's Small Bodies Node:
http://www.psi.edu/pds/archive/occ.html.} is not straightforward.
As an example, \cite{Shevchenko06} were able to use just 57
occultation diameters for albedo determinations, among which only
18 have an estimated accuracy $<5\%$. \cite{Shevchenko06} compared their
''occultation albedos'' against the albedos derived from IRAS
observations in the thermal infrared, and against the presently
available polarimetric albedos. However, because their data set
contains occultation sizes only for asteroids larger than
$\sim$100 km, they were not able to explore any size-dependence of
systematic errors of polarimetric and thermal infrared albedos.

Asteroid occultation observations - strongly encouraged by
professional astronomers -  are supported and secured by
organizations grouping mainly volunteering amateurs, such as the
International Occultation Timing Association (IOTA) or the
European Asteroid Occultation Network (EAON), strictly cooperating          % MDB strictly and not stricly
with -- and including -- professional astronomers. Today, for
stars currently included in the predictions of asteroid
occultations (usually brighter than V=12 and listed in UCAC or
Hipparcos catalogues) the main source of prediction uncertainties
resides in the accuracy of asteroid ephemerides (as discussed in
the next section). Current uncertainties in the ephemerides of
asteroids are rarely below $\sim$0.5 arcsec (except for the
largest bodies), corresponding to a displacement of $\sim$350~km
on the Bessel plane for an object at 1~AU from the Earth. The
corresponding uncertainty on Earth has similar values, thus
requiring a large number of appropriately spaced observers to
increase the success probability of actually observing the
occultation. Furthermore, in the coming years a slight degradation
in the accuracy of occultation prediction can be expected, as a
consequence of the uncertainty on the proper motions of stars.

The Euraster network data\footnote{http://www.euraster.net/results/index.html}
show that in the year
2005, $\sim$15$\%$ of the occultations observed under a clear sky
was successful. This simple evaluation, however, does not include
any information on the real position of the observers with respect
to the predicted path on Earth, nor a distinction based on the
presence of other degrading factors, such as a small flux drop,
ephemeris uncertainty, short duration, and bad atmospheric
conditions. It would thus be difficult and time-consuming to
establish a more precise ``average'' success rate, since it
depends on details that would need a case-by-case evaluation.
\begin{figure*}[t]
  \includegraphics[width=17cm]{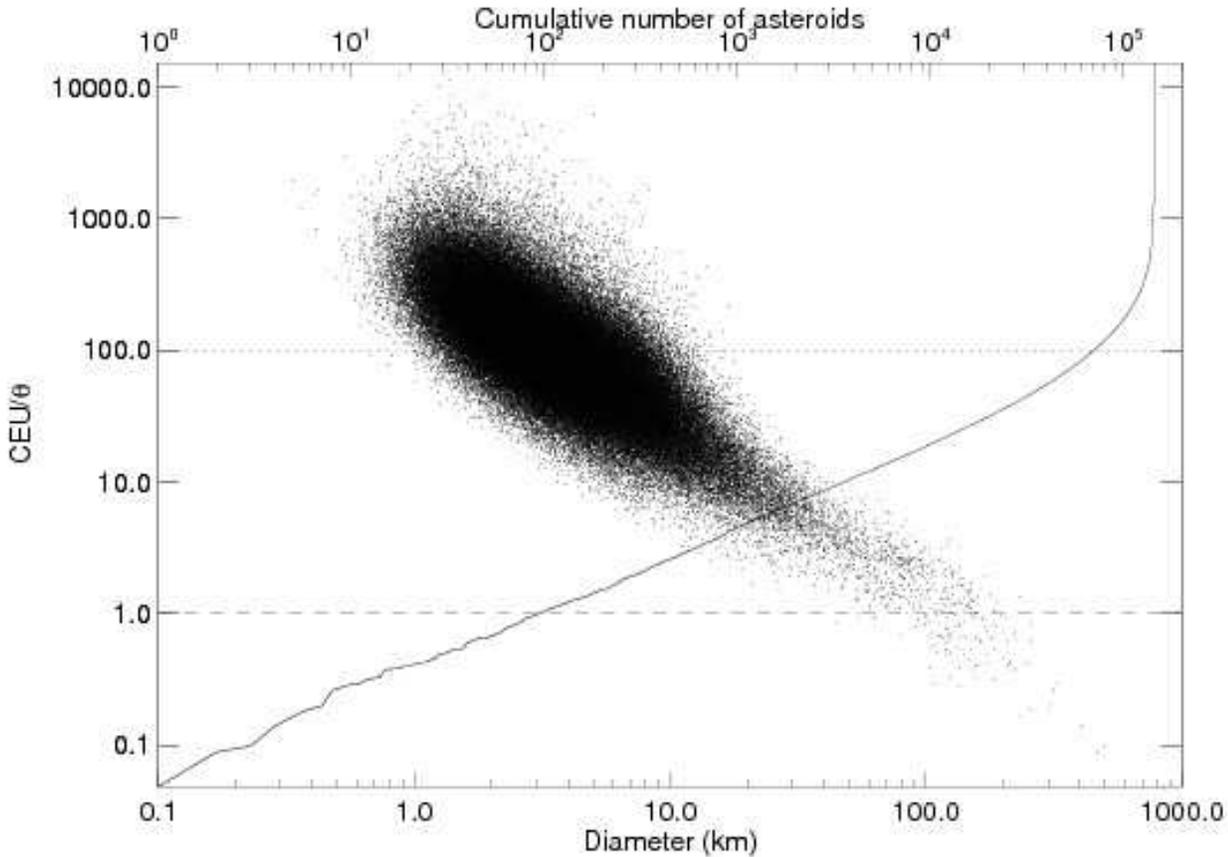}
  \caption{Plot of the ratio CEU/$\theta$ versus the object
  diameter (in km, lower X-axis)
  for the numbered main belt asteroids known and observable at the
  date of writing: \astorbdate. The solid curve, superimposed on the
  plot, shows the cumulative
  number of asteroids (upper X-axis) which have a CEU/$\theta$ ratio
  smaller or equal than the corresponding value of the Y-axis on the
  left-hand side of the figure. The curve clearly shows that there are
  about 80 - 90 asteroids with a CEU/$\theta$ value smaller than 1.
  We expect that Gaia will improve the accuracy of asteroid ephemeris
  prediction by a factor $\sim$ 100 (dotted horizontal line),
  increasing by almost three orders of magnitudes the number of
  asteroids for which CEU $\sim \theta$.
}
  \label{F:CEUtheta}
\end{figure*}

\subsection{Current accuracy of asteroid ephemeris}
\label{S:Accuracy}

The uncertainty in asteroid ephemeris can be evaluated by using
information provided by the Bowell's Asteroid Orbit Database
(ftp://ftp.lowell.edu/pub/elgb/astorb.html). The asteroid ephemeris
uncertainty in this source is calculated by means of the procedure
described by Muinonen and Bowell (1993), and it is expressed in
terms of a small set of parameters, namely the CEU (Current
Ephemeris Uncertainty), the PEU (Peak Ephemeris Uncertainty, and
its date of occurrence), and the PEU over the following 10 years.
All quantities are referred to the constantly updated epoch of the
osculating orbital elements. The method adopted for the
calculation of the CEU is considered to give meaningful estimates
of the current ephemerides uncertainty for multi-opposition Main Belt Asteroids (MBAs)
but, on the other hand, it is assumed to be unreliable for Earth crossers or asteroids
with poorly determined orbits. Since we focus our study on the
observation of occultations events of MBAs
\footnote{According to Tedesco et al. (2005) we define MBA as those
asteroids having orbits between the 4:1 and 2:1 mean-motion
resonances with Jupiter, i.e., those with orbital semimajor axes
between 2.064 and 3.278 AU and with modest inclinations
($<$25$^0$) and eccentricities ($<$0.3).}
we can assume that the CEU is representative of the
uncertainty on the position at the epoch for which the osculating
orbital elements are calculated.

Note that the CEU gives the length of the major axis of the
uncertainty ellipse (usually very slender) representing the set of
positions on the sky where the object can be found, at a 1-sigma
confidence level. This axis is usually lying on the line of
variations, or at a small angle with it. The main effect of this
orientation will thus be a major uncertainty on the event epoch,
not on the shadow path in the direction perpendicular to the
asteroid shadow motion. This latter uncertainty is the dominating
one as far as the observability of the occultation event is
concerned; in fact, a shift in the predicted epoch can easily be
compensated by an increased duration of the observation. It is
thus likely that, by retaining the CEU as an estimate for the
position uncertainty we are, in general, rather conservative. The
exception is probably represented by a small number of large
asteroids, having a smaller and more circular uncertainty ellipse,
for which the CEU correspond more strictly to the prediction
accuracy (Bowell, private communication).

Under these assumptions, we can then state that with a CEU larger
than the apparent angular size of the asteroid, occultation events
are difficult to predict. We therefore used the ratio of
the CEU to the apparent angular radius $\theta$ of the asteroid
(given by the ephemeris at the CEU epoch) as a statistical
indicator for the current predictability of occultation events.
For other epochs, of course individual CEU values will be
different, but the overall statistics remain valid.

For our computations, the apparent sizes of asteroids $\theta$ were calculated
from their IRAS diameters included in the
Supplemental IRAS Minor Planet Survey (SIMPS) of Tedesco et al.
(2002).
Note that the SIMPS contains diameters for only the first
$\sim$2200 numbered asteroids. For the large majority of
bodies in our sample, diameters are thus not known. For them, we estimated sizes from
asteroid $H$ magnitudes (which are included in the astorb.dat
file) and an assumed geometric visible albedo via Eq.~\ref{H2D}.
However, the use of a constant albedo through the entire Main Belt
would be an oversimplification: the albedo range for MBAs is very large
(0.03 - 0.5), and it is known, for instance, that an inverse
correlation between albedo and heliocentric distance exists. We
used, therefore, the procedure described by Tedesco et al. (2005)
to assign realistic albedos to the numbered asteroids not included
in the SIMPS database.

According to this model (and using the numbering as in Tedesco et al. 2005),
the main belt is divided into three zones: i.e. zone 2 for $2.064 < a < 2.50$; zone
3 for $2.50<a\leq2.82$; zone 4+5 for $2.82<a<3.278$, where $a$ is
the orbital semimajor axis of the asteroid.
To assign an albedo, we first divide MBAs according to the zone
to which they belong (2, 3 or 4+5) and to their mean apparent visible magnitude at
opposition, given by $V(a,0)=H+5\ log(a(a-1))$. Either the observed
or the bias-corrected albedo distribution of
\cite{SAMI} (Table 3) is assigned to an asteroid, depending on
whether the value of $V(a,0)$, is $<15.75$ or $\geq15.75$,
respectively.
The same table gives the zonal distribution for four
different classes of albedos: i.e. low ($0.020<p_V\leq0.089$),
intermediate ($0.089<p_V\leq0.112$), moderate
($0.112<p_V\leq0.355$), and high ($0.355<p_V\leq0.526$).
We then weight the number of asteroids according to the corresponding
zonal albedo distribution and randomly distribute albedos
uniformly within each of the albedo classes.

Fig.~\ref{F:CEUtheta} shows the resulting ratio CEU/$\theta$ for all
numbered MBAs known and observable at the date of writing
(astorb.dat file of \astorbdate) as a function of their diameter.
We assumed that an asteroid is ``observable'' (in a broad sense) if
it has a solar elongation $\geq$50$^o$.
The distribution shows that only
very few asteroids (less than 90) -- mainly bright MBAs with
diameters larger than 50 km -- have a ratio CEU/$\theta$ smaller
than one. This roughly corresponds to the number of asteroids for
which we can expect today to make reliable predictions of
occultation events, or stated in another way, their occultations
have a high probability of being observed inside the predicted
shadow path. An improvement of occultation predictions is at
present still possible through last-minute astrometry, aiming to
constrain the trajectory of the asteroid with respect to the
target star in the hours immediately preceding the event.

It is probable that only by the forthcoming all-sky astronomical surveys,
which are capable of providing accurate multi-epoch astrometry of asteroids,
will we be able to extend occultation studies of minor bodies
to much smaller sizes, with the possibility of deriving
occultation diameters for several tens of thousand of asteroids. The
all-sky survey program that will have the best ever astrometric
accuracy is the ESA mission Gaia, foreseen to be launched at the end of
2011.

However, the capability of systematic asteroid occultation
observing campaigns of providing accurate size determination
can be judged only if we can estimate the
observational effort required. Thus, we will calculate in the following
section the theoretical asteroid occultation
rate for a given site on Earth, as a function of the asteroid
diameter. Later, ephemeris and star position uncertainties will be
taken into account to estimate the success rate of asteroid
occultation measurements considering instruments with different
performances, from those of a 20 cm telescope, up to the the case
of a one-meter telescope equipped with a state-of-the-art CCD
camera.

\section{A numerical model for computing the occultation rate}

%\subsection{Simulation framework}
In order to reproduce a realistic occultation rate for asteroids
of different sizes, we need to estimate how many stars each
asteroid occults in a unit time (e.g. one day). This is
equivalent to the number of stars contained within the area on the
sky swept by each asteroid in that time interval, and visible for a
generic observer on Earth. This area can be easily calculated
by taking the product between the apparent angular diameter of the
asteroid, 2$\theta$, and the distance on the sky between the two
positions of the body calculated with the difference of one day.
In order to calculate the density of stars falling within the area
swept by each asteroid every day, we extracted the stellar density
from the realistic galactic model used by Crosta \& Mignard (2006).
In this model the number of stars per square degree, as a function of
their magnitude and their galactic coordinates, is expressed in
terms of a set of Chebyshev polynomials, whose coefficients were
determined by a fit to the measured star count.
Note that, in practice, the number of stars occulted every day by
each asteroid is much smaller than one, so to decide whether an
occultation event take place or not, we generated a random number
uniformly distributed between zero and one, and we took as
positive events those cases where the random number was found to be
greater or equal to the number of occulted stars. If the
test was positive, the epoch of the occultation event was
generated randomly inside the time step interval.
We repeated this procedure every day over a period of 5 years, in
order to consider different apparitions for each MBA, thus including
different sky patches visited by each body.

We assumed our observer to be situated at the geocenter. This
choice is not a loss of generality, since the overall statistics
will be the same for any given site on the surface of the Earth.
However, we divided by a factor of 2 the number of occultations
obtained, to take into account the half celestial sphere visible        % MDB visible and not visibile
in the presence of a theoretical, geometric flat horizon. One should
note that this correction could be different (and probably larger)
in practice, due to the difficulty of observing events at low sky
elevations. Moreover, to calculate asteroid positions we solved
the unperturbed two-body problem: planetary perturbations
represent a level of refinement not required for our purposes. As for
asteroid diameters, the same values assigned as described in
Sect.~\ref{S:Accuracy} were adopted.
Finally we computed, for each of the simulated events, the
maximum duration of the occultation $\tau_o$ using the object
angular diameter and apparent motion at the occultation epoch, the asteroid's apparent
magnitude $m_a$, and the magnitude drop during the occultation.

\begin{figure}[h]
  \includegraphics[width=9.5cm]{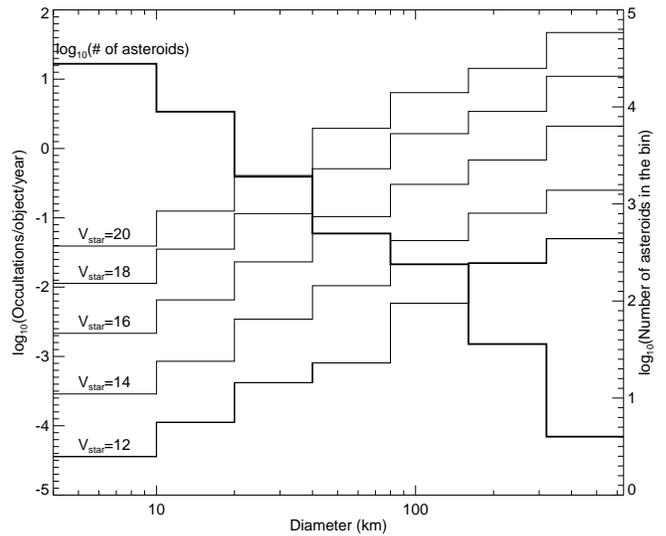}
  \caption{Number of occultations per object during one year, as a function of
the asteroid diameter (thin lines). These statistics are valid for a single site and are
computed for different values of the limiting magnitude as indicated by the
legend. The thick curve (scale on the right-hand side y axis) represents the number
of asteroids for each bin in the simulated population. The product of this last
number with each rate/object curve provides the total number of occultations per
year for a given size range.}
  \label{F:n_occ}
\end{figure}

The obtained results, of course, have a statistical value. They are
presented in Fig.~\ref{F:n_occ},
where we show the average number of occultation events that occur
for a single site and a single object, during one year.
It is clear that including faint stars strongly enhances the
occultation rate. In fact, we find that any object larger than
20-30 km occults a 20$^{th}$ magnitude star at least once per
year.

However, the theoretical occultation rate that we have computed
so far represents a theoretical value based on geometry only. It
does not take into account the ``observability'' of each
event, that is the capability of a given instrument to reveal the
temporary flux drop produced by the occultation.
Furthermore, since the scientific goal that interests us
is the impact of occultation observations on asteroid size
determination, we restricted our study to events from which the
the size of an asteroid can be determined
within a given uncertainty level.

We investigate the observability of the
events by comparing portable equipment, similar to those
currently used by most observers in present days, to a completely
different situation, i.e. that of a fixed 1-m class telescope
equipped with a cooled CCD camera. With the larger instrument,
much fainter stars (and much smaller magnitude drops) can be
measured, extending the investigation of occultation events toward
smaller asteroids. As we will see, this will in turn allow the
exploitation of improved asteroid ephemeris and stellar
astrometry.

\section{The measurement of an occultation: limits imposed by the instrument}

The observation of an occultation event consists of the measurement
of the temporal flux variation, by means of rapid photometry, of a
single source composed by the unresolved images of the star and
the asteroid. A flux drop occurs during the occultation when
only the asteroid hiding the star is observable. The duration of
the occultation is typically of a few seconds for large
(50-100~km) asteroids, and the most favorable events, of course,
involve a bright star and a faint asteroid. In this case, the flux
drop can reach several magnitudes. We will use the symbols $m_a$,
$m_s$, and $m_{sa}$ to indicate respectively the magnitude of the
asteroid, the magnitude of the star and the magnitude of the
unresolved source star + asteroid immediately before and after the
occultation. For the corresponding measured fluxes we adopt the notation
$F(m_a)$, $F(m_s)$, and $F(m_{sa})$, respectively.

To obtain useful size estimates of asteroids from the
observation of an occultation event, the use of performing
equipment, capable of exposure times considerably shorter than
the maximum event duration, and sensitive enough to allow accurate
photometric measurements of the flux drops is required. Still, a
single observation of an occultation will not yield, in general, a
direct measurement of the object's size. This is because the
measured occultation chord does not necessarily correspond to the
largest projected size of the asteroid occulting the star.
Moreover, in general, the three-dimensional shape of
the occulting asteroid and its orientation should be known for a
meaningful interpretation of the results obtained from occultation
measurements.

As for the definition of the signal to noise ration (S/N) of the photometric
measurements of the occultation event, we adopt the ratio between
the amplitude of
the flux drop ($F(m_{as}) - F(m_a)$) and the photometric uncertainty associated with
the flux during ($\sigma_{F(a)}$) and outside ($\sigma_{F(as)}$) the
occultation:
\begin{equation}
S/N = \frac{F(m_{as}) - F(m_a)}{\sqrt{\sigma_{F(a)}^2+\sigma_{F(as)}^2}}.
\end{equation}

For a given instrument setup and for each event, the available flux and the
imaging equipment characteristics determine the corresponding photometric
uncertainty.

After having chosen an instrument set (see below) we thus computed
-- for each event -- the minimum exposure time $t_m$ providing S/N$\sim$3.
%To avoid unrealistic
%situations for the few events involving extreme flux drops (i.e. very bright stars)
%$t_m$ is not allowed to be smaller than $t_m$=0.01~sec.
Assuming that the absolute timing is available for the CCD frames
with negligible uncertainty, we take $t_m$ as the estimate of the
occultation timing uncertainty, i.e. $\sigma_\tau=t_m$. Since the
maximum transit duration is proportional to the object diameter
$D$, the ratio $\sigma_D=\sigma_\tau/\tau_o$ (where $\tau_o$
represent the maximum occultation duration, on the center line)                     % MDB occulation and not occulation
will represent the relative accuracy expected for the diameter
determination.

One should note that, according to the S/N definition above, an
exposure time yielding S/N$\sim$3 will always be found, but in
some cases it will be comparable to -- or significantly longer
than -- the occultation duration.  However, we will consider
as ``non--interesting'' those events having a relative uncertainty
on the final diameter determination $\sigma_D>40\%$. By adopting
this constraint, ``unobservable'' events (i.e. those requiring
$t_m \le \tau_o$) are discarded.
We then classified the remaining events as function of $\sigma_D$.
%, regardless of the
%fact that, in general, other chords will be observed for a
%randomly chosen location.
% **PT: ricordare poi (o prima) che la generazione della simulazione degli eventi non si
% spinge fino a tirare a sorte una ``corda'' per ogni evento.
%As a consequence, one should be aware that the values of
%$\sigma_D$ are optimistic, or in other words, that different
%chords of a given object must be measured in order to really lower
%the uncertainty to the computed value.

Of course, discarded events
could still be of scientific value, for example for the search of
asteroid satellites, however in this study we do not address in detail this
otherwise interesting subject.

\begin{figure}[th]
  \includegraphics[width=9.5cm]{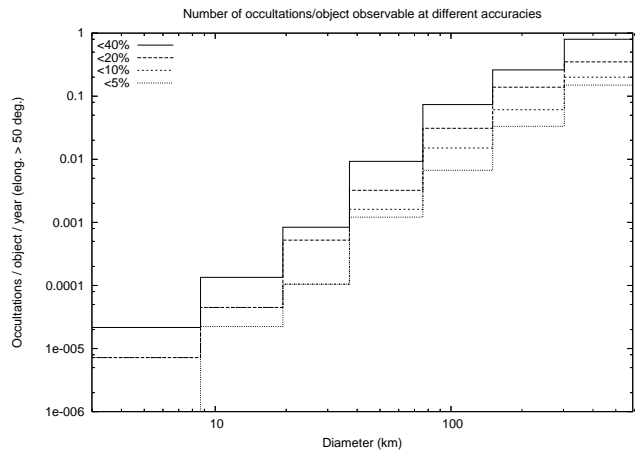}
  \caption{
Number of observable occultations per object during one year, as a
function of the asteroid diameter, for a 20 cm telescope equipped
with a cheap fast-readout camera. These statistics are valid for a
single site and are computed for different values of the maximum
relative uncertainty on size, $\sigma_D$, as indicated by the
legend. }
  \label{F:20cm}
\end{figure}

The predictions of occultation events that are currently distributed to
observers include target stars not fainter than
V=12-13, thus reducing the occultation rate at about ~0.1
events/objet/year even for the largest bodies. As a further
selection, requirements on the minimum occultation duration and
magnitude drop mean that to only the $\sim$1000
asteroids larger than $\sim$40 km are taken into account. We ran a first simulation
considering the performance of a 20~cm, f/10 telescope equipped
with a rapid, uncooled video camera (30~frames per second,
exposure 1/30~sec) with rather high readout noise
(15~e$^-$/pixel). We considered a CCD operating at 290~$^o$K, at a
pixel scale $\sim$0.9 arcsec/pixel, with a seeing FWHM around 1
arcsec. Sky background was assumed to be V=12~mag/arcsec$^2$. With
the high CCD frame rate imposed, the dominating noise source is
not dark current, but the read-out noise. The sensor temperature
is thus of secondary importance. By using aperture photometry with
an optimized aperture size, we obtain the results shown in
Fig.~\ref{F:20cm}. The integral of the occultation rate for
objects $d>40$~km yields no more than a few events observable at
high accuracy each year for a single site. This appears to be
fully consistent with the present observation statistics.

\begin{figure}[th]
  \includegraphics[width=9.5cm]{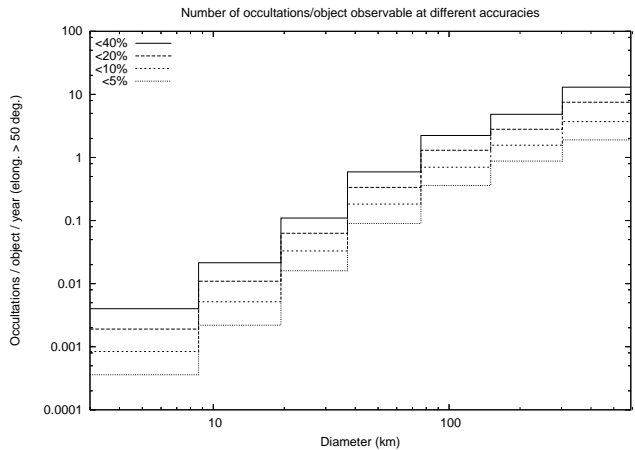}
  \caption{
Number of observable occultations per object during one year, as a
function of the asteroid diameter, for a 1~m telescope equipped
with a cooled, low-noise, fast-readout camera. The histogram is
computed, as in Fig.~\ref{F:20cm}, for different values of
$\sigma_D$. }
  \label{F:1mL3}
\end{figure}

The simulation has also been run considering a 1-m telescope
coupled to a low-noise, photon-counting CCD camera. These cameras,
based on an electron multiplier in the read-out register, operate
at very low temperatures ($\sim$180~$^o$K) and presently represent
the state--of--the--art technology for astronomical use. Considering a larger pixel
(22~$\mu m$) and a faster focal ratio (f/6), the image scale (0.76 arcsec/pixel)
is not far from that of the 20~cm telescope equipped as above.
Read-out noise is considered to be
(1~e$^-$/pixel); all other conditions are not varied relatively to
the previous numerical experiment. The increase in camera
performance and photon flux provides an improvement in the number
of events at a given size range, by a factor of 1 to 2 orders of
magnitude (Fig.~\ref{F:1mL3}).

In the following section, we use the statistics obtained with
these two extreme instrument setups as a reference, and we include in our analysis the
contribution of uncertainties on both asteroids and star
positions.

\section{Role of astrometric accuracy}

Up to now we have discussed the statistics of occultation events
and their observability as if we were able to know asteroid and
star positions with infinite accuracy. In reality, they are
affected by astrometric uncertainties. As a consequence, in
principle, an observer has to monitor a larger number of candidate              % MDB : has to and not hasto
events in order to reach the count of the positive occultations
predicted by the theoretical rate.
%
%the set of {\sl candidate} events that an observer have to monitor
%in order to observe all the occultations that actually take place
%for a given location should include all configurations that,
%although nominally predicted for other nearby sites, could
%potentially produce an occultation simply due to positional
%errors.
%
The number of events that need to be monitored can be estimated
by assuming as ``impact section'' for each asteroid the apparent size given by its
diameter plus the width of the ephemeris uncertainty at the
occultation epoch. Adding an uncertainty on the star position will
further increase the ``impact section''.

To take this effect into account, for each event we
computed an estimate of the total uncertainty on the occultation
path as: $\sigma_{path}=\sqrt{\sigma_{CEU}^2+\sigma_{*}^2}$, where
$\sigma_{CEU}$ is assumed to be equal to the CEU for the asteroid
discussed above, while $\sigma_{*}$ represents the uncertainty on
the star position. The prediction lists usually diffused to the
amateur community include the stars of the catalogues FK6 (I+III),
Hipparcos, Thyco 2, and UCAC2 for stars of magnitude 10$<$V$<$12.5
(\cite{Dunham05}). For stars with V$<$10 the UCAC2 positions, if
available, are used to replace those given in the Thyco 2. A
reasonable estimate of star position uncertainties is of about
1~mas for the FK6, 10~mas for the Hipparcos, 70~mas for Tycho 2,
and 20~mas for the UCAC2, for sources in the range of magnitudes
V$\sim$10-14, whereas the uncertainty is of about 40~mas outside
this range (Zacharias et al. 2004). For each simulated event we
thus attribute to the occulted star an uncertainty typical of
Thyco-2 or UCAC2 stars, with equal probability (this roughly
reflects the current distribution of catalogue exploitation for
predictions, see \cite{Dunham05}).

In a second step, we compute the ratio $p = 2\theta/\sigma_{path}$, 
(we recall here that $2\theta$ is the
apparent asteroid diameter) averaged this over all the events occurring
in a given diameter range. This can be considered to be a
``prediction efficiency factor''.
% I would take this out
%Since $\sigma_{path}$ is the
%standard deviation of a normal distribution,
If we assume that the uncertainties on asteroid ephemeris and star
positions are Gaussian distributed, then $\sigma_{path}$ represent
the standard deviation of the total uncertainty on the occultation
path, and $p=1$ implies a 1-sigma confidence level (68\%) of
actually being able to observe the occultation event. This could
certainly be considered as an event deserving an observation, with
high success probability. In reality, the CEU can follow a more
complex distribution, and the $\sigma_{path}$ has to be taken as a
rough estimate of dispersion rather than a true Gaussian
deviation. Thus $p$ cannot be considered more than a reasonable
guess of the success probability for each event. On the other
hand, one could also note that $1/p$, for $p<1$, is an estimate of
the number of events that should be observed to witness one
positive event, on average. For those events with p$\sim$1 the
occultation prediction is accurate and, in those cases where we
have $p>>1$ the occultation path is known so accurately that it could
be even possible to determine the position of the chord relative
to the object barycenter.

In the case of the after-Gaia statistics, different assumptions
are needed for both CEU and $\sigma_{*}$. Since the dependence of
the ephemeris accuracy on the asteroid size should preserve the
general features shown in Fig.~\ref{F:CEUtheta}, we simply assume
an improvement factor of 100 in the ephemeris uncertainty of each
object, i.e. a post-Gaia value of the CEU reduced by a factor of
100 with respect to the present one. This is a reasonable
assumption in agreement with different analytical estimates and
simulations of the improvements
in asteroid orbital elements that Gaia will allow
(\cite{Virtanen04, Muinonen04, HestroBerthier04}).
We also assume that the end-of-mission
accuracy on star positions (in mas) can be represented by:
% VERIFICARE
\begin{equation}
\sigma_{*} (mas) = 4\times10^{-2}\ e^{0.44(V-20)+1.35},\ \ V>12
\end{equation}
where V is the star magnitude. A value $\sigma_{*}=17~\mu as$ is
obtained at V=15\footnote{This accuracy is obtained from simulations
of the on-board detection (\cite{Arenou03}) and from the expected mission
performances.}.
In this case $\sigma_{*}$ saturates to the value reached at V=12 for brighter stars.

\begin{figure}[t]
  \includegraphics[width=9.5cm]{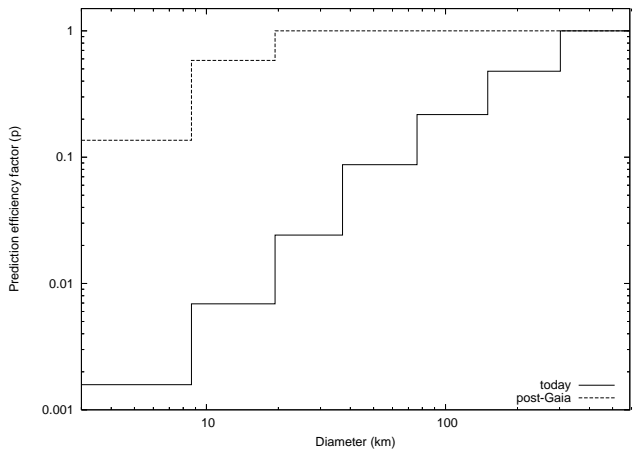}
  \caption{
Prediction efficiency factor for ``present day'' and ``post Gaia''
accuracies on asteroid ephemeris and star positions. }
  \label{F:diam_ceu_ratio}
\end{figure}

The result of the computation of $p$ is given in
Fig.~\ref{F:diam_ceu_ratio} for the present and post-Gaia
asteroid ephemerides and star position uncertainties. It is clear that 
the present occultation predictability appears
reasonable (p$>$0.1) for the largest Main Belt asteroids only (D$>$100~km),
but in the ``post-Gaia'' era an equivalent accuracy can be reached
down to $\sim$15~km.

\section{Discussion and perspectives}

Let us consider a 10-20~km asteroid as a representative example. By
multiplying the number of objects in the model at the
corresponding bin ($10^4$) with the current number of occultation
per object per year for a small instrument
(Fig.~\ref{F:20cm}), one finds that about 0.1 occultations/year
occur at a 5-10$\%$ level of size accuracy for a single site.
Furthermore, prediction efficiency (Fig.~\ref{F:diam_ceu_ratio})
is so low ($\sim$0.002) that several hundred events should be
observed (on average) to obtain one success. It is thus simply
impossible, with this equipment, to obtain accurate asteroid occultation
diameters in this size range.

On the other hand, for asteroids with $D \simeq $100~km the
predictability is rather good ($\sim$0.2) and bright star
occultations  (V$<$12), observable with the same equipment, occur
$\sim$0.01 times per object per year (at high accuracy).
Monitoring several objects by several sites will be sufficient to
obtain, over a period of years, a sample of observed positive events. This
is compatible with a lower limit at $\sim$100~km for the currently
available occultation-derived diameters.

One should note that observing occultations with larger diameters
can improve the accuracy of photometry, and increase the number of
candidate events by including smaller flux drop. However, with
present days asteroid ephemerides uncertainty, the predictability
of the occultation events remains poor, thus preventing efficient
observations at small sizes.

Much more interesting is the case of the post-Gaia scenario, in
which asteroid orbits are improved by a factor $\sim$100 and
stellar astrometry makes a revolutionary step forward for star
brightness down to $V\sim20$. In fact, in this case the prediction
efficiency remains p$=$1 (or larger) down to a diameter of 20~km. This implies
not only a perfect predictability of the events themselves, but
also an uncertainty lower than the width of the occultation shadow
on Earth. In other words, observers of occultations of asteroid
with $D>$20 km will have some indications of the position of the
chord relative to the projected shape of the object. This
represents a major change in the way asteroid occultations can be
approached, and could make them an even more powerful instrument
for asteroid studies.

\begin{figure}[t]
  \includegraphics[width=9.5cm]{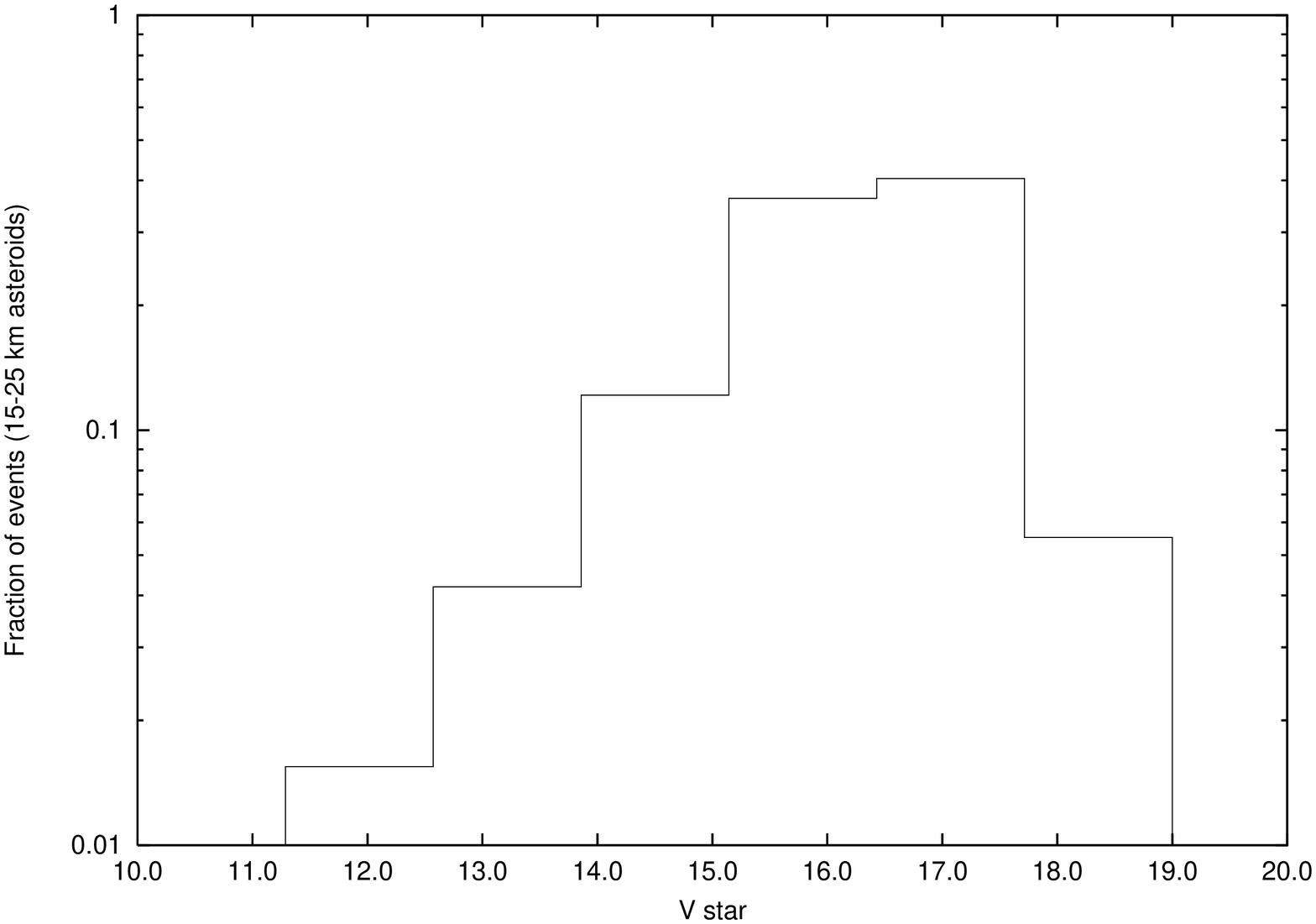}
  \includegraphics[width=9.5cm]{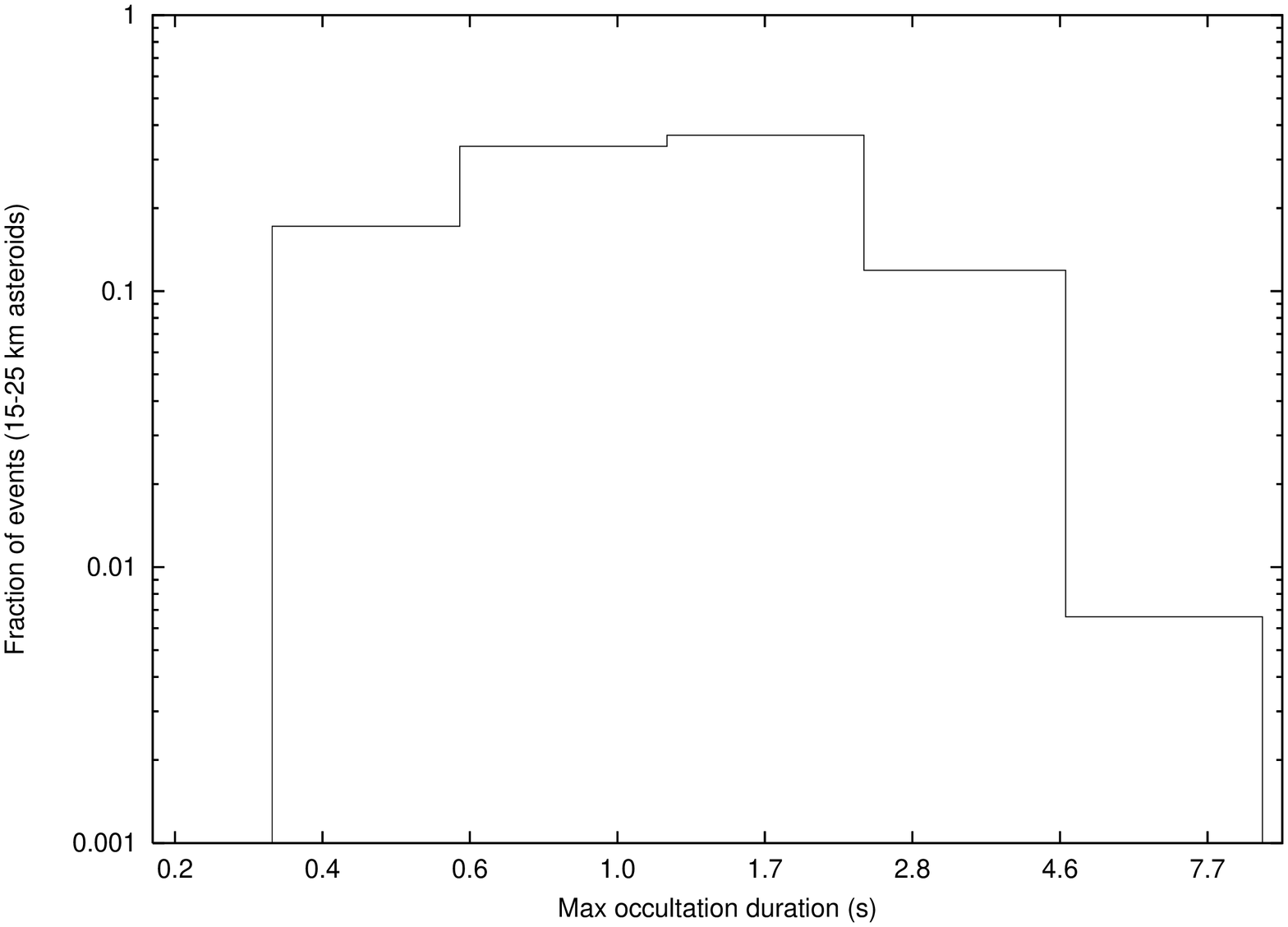}
  \caption{
Distributions of the occulted star V magnitude (upper panel) and
of the duration of the events (lower panel) for all occultations
involving asteroids in the diameter range 15-25~km.}
  \label{F:magduration20km}
\end{figure}

In Fig.~\ref{F:magduration20km} we provide the distribution of the
occultation durations and star magnitudes for events involving
asteroids in the 15-20~km range. Comfortably long occultations and
relatively bright stars are still present even toward the small
end of the asteroid size distribution for predictable
occultations. Also, we computed that $\sim$70$\%$ of the events
have a flux drop larger than 1~magnitude. However, in order to
exploit the full potential of asteroid occultations in the
post-GAIA era, more powerful instruments must be devoted to this
activity.
Figure~\ref{F:1mL3}, shows that -- by multiplying the predicted rate by the
corresponding number of objects -- 20 to 50 occultations/year are
theoretically observable with the highest success probabilities
for each 10-20~km object, by using a 1~m telescope, with diameter
uncertainties $<$10$\%$. All known objects in this size range will
lie within the CEU$/\theta\sim$~1 limit if a two-orders of
magnitude orbit improvement is reached after Gaia
(Fig.~\ref{F:CEUtheta}). This result has impressive implications,
since it means that, in principle, all objects larger than that
limit could be measured by monitoring occultations. Furthermore,
the relatively large number of events could allow a single site to
collect good statistics for each object, thus allowing an accurate
diameter determination. In practice, the large number of asteroids
involved (10$^4$) implies that a few hundred events should be
monitored each day. Of course, this is a huge effort for a single instrument
or team, but it remains feasible if spread over a few years or if
several sites with similar instruments are cooperating.

The number of events at smaller asteroid diameters will still be
interesting, although completeness will be out of reach. A
selection of objects with the lowest ephemerides uncertainty will
represent a good sample for extending size distributions toward
km-sized asteroids.

Gaia will not only provide accurate astrometry, but also
photometry and spectroscopy. Experiments of lightcurve inversion
from spare data (\cite{Cellino06}) show that shapes and rotation
vectors could be retrieved for potentially all the main belt
asteroids from Gaia observations. Earth-based surveys such as
Pan-STARRS could provide a significant contribution, although
photometric accuracy will be much lower. However, the availability
of shapes and physical ephemerides will allow extraction of a maximum
of information from occultation studies, since the orientation of
the asteroid as projected on the sky at the occultation epoch will
be known.

Several asteroid satellites will also be discovered and their
relative obits determined, thus enriching even more the potential
of further occultation events.

In conclusion, all the statistical arguments presented in this
study indicate that, when Gaia observations will be available, a
small set of dedicated 1~m telescopes will be able to provide --
within a few years -- a complete census of asteroid diameters down to
$\sim$10~km in size. The equipment foreseen to be used for future
asteroid occultation studies such as fast, low noise CCD cameras,
represent the state-of-the-art present-day technology, but it
is reasonable to predict that in $\sim$10 years from now (the after-Gaia
era) even more optimized and performing devices will be available.
Today, low noise CCD cameras are beginning to be used in
professional occultation equipment (\cite{Souza06}) and we can
expect that they will encounter a wide diffusion in the future.
Considering the constant progress made in parallel by instruments
accessible to the non-professional community, it is also highly
likely that the role of dedicated amateur astronomers will
continue to be a precious source of scientific information for
asteroid studies.

\section{Acknowledgments}
\label{Ackno}

The work of Marco Delbo was supported by the European Space Agency
(ESA External Fellowship Program). A special thank goes to F.~Mignard for
having provided the star density model, and for the always stimulating discussions.

\end{document}